\documentclass[aps,amssymb, amsmath,nobibnotes, floatfix]{revtex4-2}
\usepackage[utf8]{inputenc}
\usepackage{bm}%
\usepackage[colorlinks=true,linkcolor=blue]{hyperref}%
\expandafter\ifx\csname package@font\endcsname\relax\else
 \expandafter\expandafter
 \expandafter\usepackage
 \expandafter\expandafter
 \expandafter{\csname package@font\endcsname}%
\fi
\usepackage{xcolor}
\usepackage{verbatim}
\usepackage{graphicx}

\begin{document}

\title{Modulational instability of nonuniformly damped, broad--banded waves: applications to waves in sea--ice}
\author{Raphael Stuhlmeier}
\email{raphael.stuhlmeier@plymouth.ac.uk}
\author{Conor Heffernan}
\affiliation{School of Engineering, Computing \& Mathematics, University of Plymouth, PL4 8AA, Plymouth, UK}
\author{Alberto Alberello}
\author{Emilian P\u{a}r\u{a}u}
\affiliation{School of Mathematics, University of East Anglia, NR4 7TJ, Norwich, UK}

\newcommand{\bk}{\mathbf{k}}
\newcommand{\T}{T_{aabc}}
\newcommand{\ca}{c_{g,a}}
\newcommand{\cb}{c_{g,b}}
\newcommand{\cc}{c_{g,c}}
\newcommand{\ud}{\text{d}}

\begin{abstract}
This paper sets out to explore the modulational (or Benjamin-Feir) instability of a monochromatic wave propagating in the presence of damping such as that induced by sea-ice on the ocean surface. The fundamental wave motion is modelled using the spatial Zakharov equation, to which either uniform or non-uniform (frequency dependent) damping is added. By means of mode truncation the spatial analogue of the classical Benjamin-Feir instability can be studied analytically using dynamical systems techniques. The formulation readily yields the free surface envelope, giving insight into the physical implications of damping on the modulational instability. The evolution of an initially unstable mode is also studied numerically by integrating the damped, spatial Zakharov equation, in order to complement the analytical theory. This sheds light on the effects of damping on spectral broadening arising from this instability. 
\end{abstract}

\maketitle

\section{Introduction}
\label{sec:Introduction}

Water waves on the open sea are typically characterised by their permanence. This remarkable fact can be attested by any surfer on Hawai'i's North Shore waiting for the swell from an Alaskan storm. Attenuation due to viscous effects is sufficiently minor that the starting point of the vast majority of studies on water waves are the inviscid, incompressible Euler equations. These equations give rise to a variety of PDE model equations, many of them possessing extraordinary mathematical structure \cite{Johnson:2003ul}, much of which has been elucidated only in recent decades. 

The history of the water wave problem goes back more than two centuries, and the first approaches to the problem assumed small wave amplitudes and an essentially linearised system of equations. These equations can be seen to have solutions in the form of monochromatic waves, which consist of a single Fourier mode. Throughout the course of the 19th and early 20th centuries periodic, travelling waves of permanent form (sometimes called Stokes waves) were shown to exist mathematically, not just within the linear problem, but also for a series of ever more general formulations of the governing equations (see \cite{Toland1996} and references therein).

Despite these mathematical existence proofs, which go back to work by Levi-Civita \cite{Levi-Civita1925}, it is remarkably difficult to generate monochromatic waves experimentally, or to observe them in nature. One of the principle reasons for this difficulty was established by T.~B.~Benjamin and J.~E.~Feir \cite{Benjamin1967a}: monochromatic waves are unstable to small disturbances for a rather large range of relevant parameters. Once generated in a wave flume, one Fourier mode transfers energy to its neighbours in a so-called degenerate quartet interaction, provided the modes satisfy (at least up to small disturbances) a resonance condition. This so-called modulational instability is ubiquitous in many natural contexts \cite{vanderhaegen2021extraordinary}, and in water waves is commonly referred to as the Benjamin-Feir instability.

This raises an obvious question: if water waves are unstable and difficult to generate, how do surfers catch the swell from storms that occurred thousands of kilometres away, and how do oceanographers track these storms across entire ocean basins \cite{Snodgrass1966}? In a series of papers Segur, Henderson and collaborators \cite{Segur2005,Henderson2012} studied this phenomenon, and showed that even small amounts of damping can stabilise the Benjamin-Feir instability. Thus, while it seems that damping has a negligible effect on the propagation of waves per se, it can play a subtle yet critical role in governing wave instabilities on the open sea. In other physically important settings damping plays a more immediately visible role -- such is the case when waves propagate into sea-ice, which case provides the impetus for the present study. A noteworthy feature of damping due to sea-ice is its nonuniform (frequency-dependent) nature, see \cite{meylan2018jgr}.

Segur and collaborators considered the Benjamin-Feir instability from the perspective of a nonlinear Schr\"odinger equation (NLS) with uniform damping. Such dissipative NLS equations go back to early work of Lake et al \cite{Lake1977} (see also the more rigorous recent derivation by Dias et al \cite{Dias2008}).  As the NLS has restrictions on spectral bandwidth, attempts have been made to extend this formulation by considering a uniformly damped Dysthe (or higher-order NLS) equation. Due to the mathematical similarities between damping and forcing, studies of either phenomenon are generally complementary. Thus we find early work on the initial growth as well as long time evolution of narrowbanded surface waves under wind forcing by Hara \& Mei \cite{Hara1991}, who derived a Dysthe equation with forcing. Subsequent studies by Carter \& Govan \cite{Carter2016} and Armaroli et al \cite{Armaroli2018} likewise focused on forced and damped Dysthe-equations.

In the present work we shall use the Zakharov equation as our point of departure. This has the advantage that NLS, Dysthe and modified Dysthe equations can be derived directly from it \cite{Stiassnie1984c}, so that it generalises previous work. In addition, we shall use a spatial formulation of this equation due to Shemer and co-workers \cite{Shemer2001}, which will allow us to naturally consider the propagation of waves into sea-ice (or any other medium that can be modelled as nonuniform dissipation) which occupies a defined spatial domain.

In Section \ref{sec:Background} we provide some background for the temporal Zakharov equation, and introduce the allied spatial equation with damping. We demonstrate how this can be used to derive the damped, spatial nonlinear Schr\"odinger equation. Subsequently we restrict our attention to the three modes involved in modulational instability, and in Section \ref{sec:A dynamical system for uniform damping} analyse the simpler case of uniform damping from a dynamical systems perspective. The more realistic case of frequency dependent damping, where the symmetry between the side bands is broken, is explored in Section \ref{sec:Dynamical system for nonuniform damping}. We apply the foregoing theory to some examples in Section \ref{sec:Modelling of damped waves}, and also go beyond the classical modulational instability to explore the effect of damping on spectral broadening. Finally we provide some concluding remarks and perspectives for future work in Section \ref{sec:Discussion}.

\section{Background}
\label{sec:Background}
\subsection{The temporal Zakharov equation}
\label{ssec:Temporal ZE}

In the inviscid, incompressible water wave problem, the reduced Hamiltonian formulation due to Zakharov \cite{Zakharov1968} and Krasitskii \cite{Krasitskii1994} captures resonant and near-resonant interactions which occur with cubic nonlinearity in deep water. This Hamiltonian is \cite[Eq.\ (2.22)]{Krasitskii1994}
\begin{equation*}
    H = \int \omega_0 b_0^* b_0 d k_0 + \frac{1}{2} \int T_{0123} b_0^* b_1^* b_2 b_3 \delta_{0+1-2-3} \ud k_{0}\ud k_1 \ud k_2 \ud k_3,
\end{equation*}
where $b_i=b(k_i,t)$ are canonical variables related to the Fourier transforms of the free surface and potential at the free surface, $\omega_0$ is the linear frequency, $T_{0123}=T(k_0,k_1,k_2,k_3)$ is the interaction kernel and $\delta$ the Dirac delta distribution. Subscripts are used to denote wavenumber, so that, for example, $\omega_0 = \omega(k_0).$ 

The corresponding equation of motion, called the Zakharov equation after \cite{Zakharov1968} is 
\begin{equation} \label{eq:tZE}
    i \frac{\partial b_0}{\partial t} = \frac{\delta H}{\delta b_0^*} = \omega_0 b_0 + \int T_{0123} b_1^* b_2 b_3 \delta_{0+1-2-3} \ud k_{1}\ud k_2\ud k_3.
\end{equation}
The version of the Zakharov equation shown here is restricted to third-order in nonlinearity, and is derived by suitable elimination of non-resonant terms. It gives rise to a number of useful model equations, chief among them the nonlinear Schr\"odinger (NLS) family of equations, which we shall explore in greater depth below. In fact, the form of the equation \eqref{eq:tZE} is generic in any dispersive medium where four-wave interactions are allowed \cite{Zakharov1992a}. The physics of the water wave problem are encoded entirely in the kernel function $T$.  More details and further references to the Zakharov equation in the context of water waves can be found in the recent review  \cite{Stuhlmeier2024}.

\subsection{The Spatial Zakharov equation and inclusion of damping}
\label{ssec:Spatial ZE & Damping}
While the Zakharov equation is an integro-differential equation for time-evolution, having undertaken a Fourier transform in space, in many situations of practical interest it is necessary to consider the converse situation. Spatial evolution of waves is the situation encountered in flume experiments, and is also the appropriate viewpoint for waves propagating into a confined region with significant dissipation, e.g.\ an area of sea-ice. In such cases the temporal Zakharov equation \eqref{eq:tZE} must be replaced by a corresponding spatial evolution equation first derived by Shemer and co-workers \cite{Shemer2001}. It is written
\begin{align}\nonumber
i c_g \frac{\partial B(x,\omega)}{\partial x} = &\iiint T(k,k_1,k_{2},k_{3}) B^*(x,\omega_1) B(x,\omega_2) B(x,\omega_3) \\
&\cdot  \exp(-i(k+k_1-k_2-k_3)x) \delta(\omega+\omega_1 - \omega_2 - \omega_3) \ud \omega_1 \ud \omega_2 \ud \omega_3.
\label{eq:SpatialZE}
\end{align}
in one spatial dimension. The main difference between the spatial and temporal cases is the appearance of a group velocity coefficient $c_g$ in the former as well as the (near) resonance condition now being expressed as $\omega + \omega_1 - \omega_2 - \omega_3 = 0$ with $|k+k_1-k_2-k_3| = \mathcal{O}(\epsilon^2)$ where $\epsilon$ is the wave steepness \cite{Kit2002}. 

This equation can be transformed into autonomous form by writing $B_i = b_i \exp(-ik_i x):$
\begin{equation}
\label{eq:Spatial ZE Discr2}
c_{g,j} \left( i\frac{d b_j(x)}{dx} + k_j b_j \right) = \iiint T_{jlmn} b_l^* b_m b_n  \delta(\omega_j + \omega_l - \omega_m - \omega_n) \ud \omega_l \ud\omega_m \ud\omega_n.
\end{equation}
It is important to recall that subscripts in \eqref{eq:Spatial ZE Discr2} now denote frequency rather than wavenumber, although the kernel $T_{jlmn}$ remains $T(k(\omega_j),k(\omega_l),k(\omega_m),k(\omega_n)).$

Adding a spatial damping term can be accomplished by writing $(k_j+i\gamma_j)$ in place of $k_j$ in the second term on the left-hand side. This is analogous to the inclusion of damping by modifying the frequency in the temporal Zakharov equation \eqref{eq:tZE}, see Shrira et al \cite{Shrira1996}. In our study, we shall consider both the simpler case of uniform damping, as well as the more general case of frequency dependent damping, which typically occurs when waves propagate in sea ice. 

The particular frequency-dependence of the damping depends strongly on characteristics of the medium itself, and analytical as well as experimental work suggests a sea-ice damping of the form 
\begin{equation} \label{eq:damping-form} \gamma = s \times \omega^n,\end{equation}
which is the form of damping parameter we will employ in what follows. Uniform damping can be achieved by using the parameter $\gamma$ corresponding to the carrier mode for all other modes. Alternatively, each mode $\omega_i$ may have a distinct damping coefficient $\gamma_i$ for a given value of $s.$ The majority of cases we discuss will employ a moderate damping $s=O(10^{-6}),$ and we will take $n=3$ throughout, as described in \cite{Alberello2023}.

\subsection{Derivation of the damped spatial NLS}
\label{ssec:Derivation of dNLS}

The Zakharov equation is a powerful yet underutilised tool for studying the evolution of water waves, particularly the present damped, spatial formulation. In order to put it into the context of a larger literature focused on the damped spatial nonlinear Schr\"{o}dinger equation (see \cite{Alberello2023,Alberello2022,Kimmoun2016}) we demonstrate how the latter can be derived.

The central assumption needed to derive the NLS is that all interacting waves are clustered about a single wavenumber, say $\bk_0,$ an assumption usually referred to as ``narrow-bandwidth". Such an assumption can be imposed when deriving the equation via perturbation theory \cite{Johnson1997}, as well as when starting from the Zakharov formulation. In the latter case, the Zakharov kernel is replaced by the kernel $T(\bk_0,\bk_0,\bk_0,\bk_0)$, and the frequency $\omega(\bk)$ is expanded in a Taylor series about $\omega(\bk_0).$ These two steps allow the inverse Fourier transform to be carried out, and lead to the NLS in much the same way that Zakharov \cite[Eq.\ (2.7)ff]{Zakharov1968} first outlined. Our derivation below closely follows Kit \& Shemer's \cite{Kit2002} derivation of the Dysthe equation, but includes a damping term. 

Because the NLS is an equation for the free surface envelope, we first relate the complex amplitudes $B$ of the Zakharov formulation to the free surface $\zeta$ at lowest order (i.e.\ without bound modes) by
\begin{equation}
\label{eq:Free Surface}
    \zeta(x,t) = \frac{1}{2\pi}\int_{-\infty}^{\infty} \left(\frac{\omega}{2g}\right)^{\frac{1}{2}} \left[B(\omega,x) \exp(i (k(\omega)x - \omega t)) + c.c.\right] \ud\omega, 
\end{equation}
where $c.c.$ denotes the complex conjugate of the preceding expression, and we write $k(\omega)$ for clarity, invoking the linear dispersion relation. 

As a first step we write all frequencies $\omega_i$ in terms of a central (carrier) frequency $\omega_0$ and a small perturbation $\omega_i'$ with ${\omega_i'}\ll {\omega_0}$, i.e.\ $\omega_i = \omega_0 + \omega_i'.$ Introducing a new variable $A$ 
\begin{align*}
    A(\omega',x) &= B(\omega,x)\exp(i(k(\omega_0 +\omega') - k_0)x),
\end{align*}
and substituting into \eqref{eq:Spatial ZE Discr2} yields
\begin{align*}
    i\frac{\partial A}{\partial x} + \left( k(\omega_0 + \omega') - k_0 \right) A + i\gamma A 
    = \frac{k^3_0}{2g\pi^2}(\omega_0+\omega')\int_{\mathbb{R}^3}\, A^{*}(\omega_1',x)A(\omega_2',x)A(\omega_3',x) \delta_{0,1}^{2,3} \ud \mathbf{\omega'},
\end{align*}
where $T = T(\omega_0+\omega',\omega_0+\omega_1',\omega_0+\omega_2',\omega_0+\omega_3')\sim k_0^3/4 \pi^2$, $\ud \omega' = \ud \omega_1' \ud \omega_2' \ud \omega_3'$ and $\delta_{0,1}^{2,3}=\delta(\omega'+\omega'_1-\omega'_2-\omega'_3)$ . The  group velocity $c_g$ can be expanded in terms of the small perturbation as 
\begin{align*}
    k(\omega_0+\omega')-k_0 &= \frac{2k_0}{\omega_0}\omega'  + \frac{k_0}{\omega_0^2}\omega'^2 + \mathcal{O}(\epsilon^4),
\end{align*}
and using the deep water dispersion relation $c_g^{-1} = 2\sqrt{g k}/g$ yields
\begin{equation} 
\label{eq:NLS-eq-A}
    i\frac{\partial A}{\partial x} + \left( k(\omega_0 + \omega') - k_0 \right) A + i\gamma A = \frac{k_0^3}{g\pi^2}(\omega_0+\omega')\int_{\mathbb{R}^3} A^{*}_1A_2A_3 \delta_{0,1}^{2,3} \,\ud \mathbf{\omega'},
\end{equation}
where $A_j=A(\omega'_j,x)$. Now, the free surface $\eta(x,t)$ can be related to an envelope amplitude $a(x,t)$ through 
\begin{equation*}
    \eta(x,t) = a(x,t)\exp(i(k_0x-\omega_0 t)),
\end{equation*}
and the relation between $A$ and the complex amplitude $a$ is given as
\begin{align*}
    a(x,t) &= \frac{1}{2\pi}\left(\frac{2\omega_0}{g}\right)^{\frac{1}{2}}\int_{-\infty}^{\infty} \left(\left(1+\frac{\omega'}{2\omega_0}\right)A(\omega',x) \exp(-i\omega' t)\right) \ud \omega'.
\end{align*}
where the factor $1+\omega'/2\omega_0$ comes from expansion of $\sqrt{\omega/(2g)}.$

Multiplying \eqref{eq:NLS-eq-A} by a factor of $\sqrt{2\omega_0/g}(1+\omega'/2\omega_0)$ and taking the inverse Fourier transform yields the left-hand side \[ i\left(a_x + \frac{2k_0}{\omega_0}a_t\right) - \frac{k_0}{\omega_0^2}a_{tt} +i\gamma a.\]
The right-hand side is handled using the substitution 
\begin{equation*}
    \left(1 + \frac{\omega_2' + \omega_3' - \omega_1'}{2\omega_0}\right) \approx \left(1 + \frac{\omega_2'}{2\omega_0}\right) \left(1 + \frac{\omega_3'}{2\omega_0}\right)\left(1 - \frac{\omega_1'}{2\omega_0}\right),
\end{equation*}
which gives the term $k_0^3  a |a|^2.$

The equation can be rewritten in dimensionless form and in a moving coordinate frame by introducing new variables
$$a= a_0\psi, \quad \epsilon\omega_0\left(\frac{2k_0}{\omega_0}x - t\right) = \tau, \quad \epsilon^2k_0 x = \zeta, \quad \gamma = \epsilon^2k_0 \Gamma,$$
where $\epsilon = a_0k_0$ is the wave-steepness for a wave of amplitude $a_0$ and wavenumber $k_0$ as
\begin{equation}
\label{eq:damped-NLS}
     \psi_{\zeta} + i \psi_{\tau\tau} + i|\psi|^2 \psi = -\Gamma \psi.
\end{equation}
This is the form of the NLS used in numerous studies of damped and forced water waves. The classical theory of modulational instability can then be approached by inserting a monochromatic wave into \eqref{eq:damped-NLS} and performing a linear stability analysis by perturbing this wave with two small ``sidebands".

\subsection{Discretisation and the damped spatial Zakharov equation}
\label{ssec:Discretisation and damped sZE}

Our aim is primarily an analytical study of the damped, spatial Benjamin-Feir instability without the restrictions made in the preceding section, which lead to the NLS. To study this instability, which arises from the interaction of three frequencies, we first discretise the autonomous, spatial Zakharov equation \eqref{eq:Spatial ZE Discr2} with damping by substituting $ B = \sum_i B_i \delta(\omega-\omega_i)$, yielding 
\begin{equation}
 i\frac{d b_j(x)}{dx} + (k_j+i\gamma_j) b_j  = \frac{1}{c_{g,j}}\sum_{l,m,n} T_{jlmn} b_l^* b_m b_n  \delta(\omega_j + \omega_l - \omega_m - \omega_n). \label{eq:discrete damp}
\end{equation} 

The resonant set of waves which seeds the modulational instability is one which satisfies the degenerate resonance condition $2\omega_a = \omega_b + \omega_c$, where we will interpret $\omega_a$ as the frequency of the carrier and $\omega_b, \, \omega_c$ as the two side bands. This restriction results in a system of coupled differential equations for the three complex amplitudes:
\begin{align*}
    i \frac{db_a}{dx} + b_a (k_a + i \gamma_a) &= \frac{1}{c_{g,a}}\left( T_a |b_a|^2 b_a + 2 \sum_{j\neq a} T_{aj} |b_j|^2 b_a + 2 T_{aabc} b_a^* b_b b_c \right) \\
    i \frac{db_b}{dx} + b_b (k_b + i \gamma_b) &= \frac{1}{c_{g,b}}\left( T_b |b_b|^2 b_b + 2 \sum_{j\neq b} T_{bj} |b_j|^2 b_b +  T_{aabc} b_c^* b_a b_a \right) \\
    i \frac{db_c}{dx} + b_c (k_c + i \gamma_c) &= \frac{1}{c_{g,c}}\left( T_c |b_c|^2 b_c + 2 \sum_{j\neq c} T_{cj} |b_j|^2 b_c + T_{aabc} b_b^* b_a b_a  \right)
\end{align*}
Here we have used the abbreviation $T_i$ for the self-interaction kernel $T_{iiii}$, and the abbreviation $T_{ij}$ for the two-wave interaction kernel $T_{ijij},$ noting that due to symmetry $T_{ij}=T_{ji}$ \cite{Krasitskii1994}.

It is convenient to separate the real and imaginary parts of the above system by writing $b_i = \sqrt{I_i}\exp(i \phi_i)$ where $I_i:=|b_i|^2:$ 
\begin{subequations}
\begin{align} \label{eq:dIadx}
    c_{g,a}\frac{dI_a}{dx} &= -2c_{g,a}\gamma_aI_a - 4T_{aabc}I_a\sqrt{I_b}\sqrt{I_c}\sin(\theta_{}), \\
    c_{g,b}\frac{dI_b}{dx} &= -2c_{g,b}\gamma_bI_b + 2T_{aabc}I_a\sqrt{I_b}\sqrt{I_c}\sin(\theta_{}), \\
    c_{g,c}\frac{dI_c}{dx} &= -2c_{g,c}\gamma_cI_c + 2T_{aabc}I_a\sqrt{I_b}\sqrt{I_c}\sin(\theta_{}), \\
    2\frac{d\phi_a}{dx} &= 2k_a - \frac{2}{c_{g,a}}T_aI_a -\frac{4}{c_{g,a}}T_{ab}I_b - \frac{4}{c_{g,a}}T_{ac}I_c - \frac{4}{c_{g,a}}T_{aabc}\sqrt{I_b}\sqrt{I_c}\cos(\theta_{}), \label{eq:phi1}\\
    -\frac{d\phi_b}{dx} &= -k_b + \frac{1}{c_{g,b}}T_{b}I_b +\frac{2}{c_{g,b}}T_{ab}I_a + \frac{2}{c_{g,b}}T_{bc}I_c + \frac{1}{c_{g,b}}T_{aabc}\frac{I_a\sqrt{I_c}}{\sqrt{I_b}}\cos(\theta_{}), \label{eq:phi2}\\
    -\frac{d\phi_c}{dx} &= -k_c + \frac{1}{c_{g,c}}T_{c}I_c +\frac{2}{c_{g,c}}T_{ac}I_a + \frac{2}{c_{g,c}}T_{bc}I_b + \frac{1}{c_{g,c}}T_{aabc}\frac{I_a\sqrt{I_b}}{\sqrt{I_c}}\cos(\theta_{}).\label{eq:phi3}
\end{align}
\end{subequations}
Here we have defined $\theta_{} := 2\phi_a-\phi_b-\phi_c.$ 
While the presence of the damping term means that we lack conservation of energy (see Krasitskii \cite{Krasitskii1994} or Andrade \& Stuhlmeier \cite{Andrade2023instability}), there are nevertheless simplifications to be made which provide insight into the behaviour of this system. Chief among these is the identification of the \textit{dynamic phase} $\theta_{}$ as the sole phase variable of interest in the problem. In the next section we shall find the simplest three-dimensional dynamical system which encapsulates the dynamics of the spatial Benjamin-Feir instability in the case of uniform damping. 

\section{A dynamical system for uniform damping}
\label{sec:A dynamical system for uniform damping}

In analogy with the temporal evolution of conservative systems of three and four waves, as studied by Capellini \& Trillo \cite{Cappellini1991} or Andrade \& Stuhlmeier \cite{Andrade2023,Andrade2023instability}, we now aim to introduce new variables to reduce the dimension of the system \eqref{eq:dIadx}--\eqref{eq:phi3}.

We define the following two quantities:
\begin{align} \label{eq:def wave action}
    A &= c_{g,a}I_a + c_{g,b}I_b + c_{g,c}I_c,\\ \label{eq:def sideband fraction}
    \alpha &= \frac{c_{g,b}I_b}{A} - \frac{c_{g,c}I_c}{A}.
\end{align}
$A$ is akin to the wave action, a measure of total energy, while $\alpha$ is the side-band energy fraction.

Under the assumption of uniform damping $\gamma_a =\gamma_b=\gamma_c=\gamma$, we easily find 
\begin{align} \label{eq:A' uniform damping}
    \frac{dA}{dx} &= -2\gamma A,\\
    \frac{d\alpha}{dx} &= 0.
\end{align}
Thus $A$ (which we think of as a measure of the energy of the three-wave system) decreases monotonically, while the side-band energy fraction $\alpha$ remains constant.

The substitution of 
\begin{align}
    I_a &= \frac{A}{c_{g,a}}\eta, \label{eq:singlevar1}\\
    I_b &= \frac{A}{2 c_{g,b}} (1-\eta), \label{eq:singlevar2}\\
    I_c &= \frac{A}{2 c_{g,c}} (1-\eta ),\label{eq:singlevar3}
\end{align}
into \eqref{eq:dIadx}--\eqref{eq:phi3}, in which $\eta$ is 
 the energy exchange parameter,  results in the pair of evolution equations
\begin{align} \label{eq:eta' uniform damping}
    \frac{d\eta}{dx} &= -A\Omega_2\eta(1-\eta)\sin(\theta),\\ \label{eq:theta' uniform damping}
    \frac{d\theta}{dx} &= \Delta + A\Omega_0 + A\Omega_1\eta -A\Omega_2(1-2\eta)\cos(\theta),
\end{align}
in the dynamic phase $\theta$ and energy exchange parameter $\eta,$ where
\begin{align*}
   \Omega_0 &= \frac{1}{2}(\bar{T}_{c}+\bar{T}_b) + 2 (\bar{T}_{bc} - \bar{T}_{ab} - \bar{T}_{ac}), \\
   \Omega_1 &= 4(\bar{T}_{ab}+\bar{T}_{ac}) - 2\left(\bar{T}_a + \bar{T}_{bc} + \frac{1}{4} \left(\bar{T}_b +\bar{T}_c\right)\right),\\
   \Omega_2 &= \frac{2 T_{aabc}}{c_{g,a}\sqrt{c_{g,b}c_{g,c}}},
\end{align*}
and we denote $\bar{T}_{ij} = T_{ij}/c_{g,i}c_{g,j}$ and $\Delta=2k_a-k_b-k_c$ for brevity. Note that when $\eta=0$ only modes $\omega_b$ and $\omega_c$ are present, while when $\eta=1$ we have the monochromatic wave $\omega_a$ only. Together, this gives a rather simple dynamical system for the interaction of three waves with uniform damping consisting of equations \eqref{eq:A' uniform damping}, \eqref{eq:eta' uniform damping}--\eqref{eq:theta' uniform damping}, defined in the three dimensional phase space $\{ (A,\theta,\eta) \mid A\in \mathbb{R}_{\geq 0}, \, \theta \in [-\pi,\pi], \, \eta \in [0,1] \}.$

\subsection{Phase plane analysis}
\label{ssec:UD Phase plane analysis}

The undamped system \eqref{eq:eta' uniform damping}--\eqref{eq:theta' uniform damping} with $\gamma=0$ is Hamiltonian, therefore the only fixed points are saddles and centres. This setting is formally similar to the temporal evolution scenario described recently by Andrade \& Stuhlmeier \cite{Andrade2023}. Together with the simple, cylindrical phase space, this makes the dynamics rather straightforward to describe and classify.

The addition of uniform damping changes these dynamics, and the system \eqref{eq:A' uniform damping}, \eqref{eq:eta' uniform damping}--\eqref{eq:theta' uniform damping} no longer has fixed-points. Rather trajectories tend towards $A=0,$ which means that all three modes $I_a, \, I_b$ and $I_c$ decay in amplitude. One consequence of this is that separatrices connecting fixed-points of the conservative system can be traversed -- solutions otherwise confined to a restricted portion of phase space may take on other values of $\eta$ and $\theta,$ and periodic orbits around centre points of the conservative system generally lose their periodicity.

\begin{figure}[h!]
    \centering
    \includegraphics[width=\linewidth]{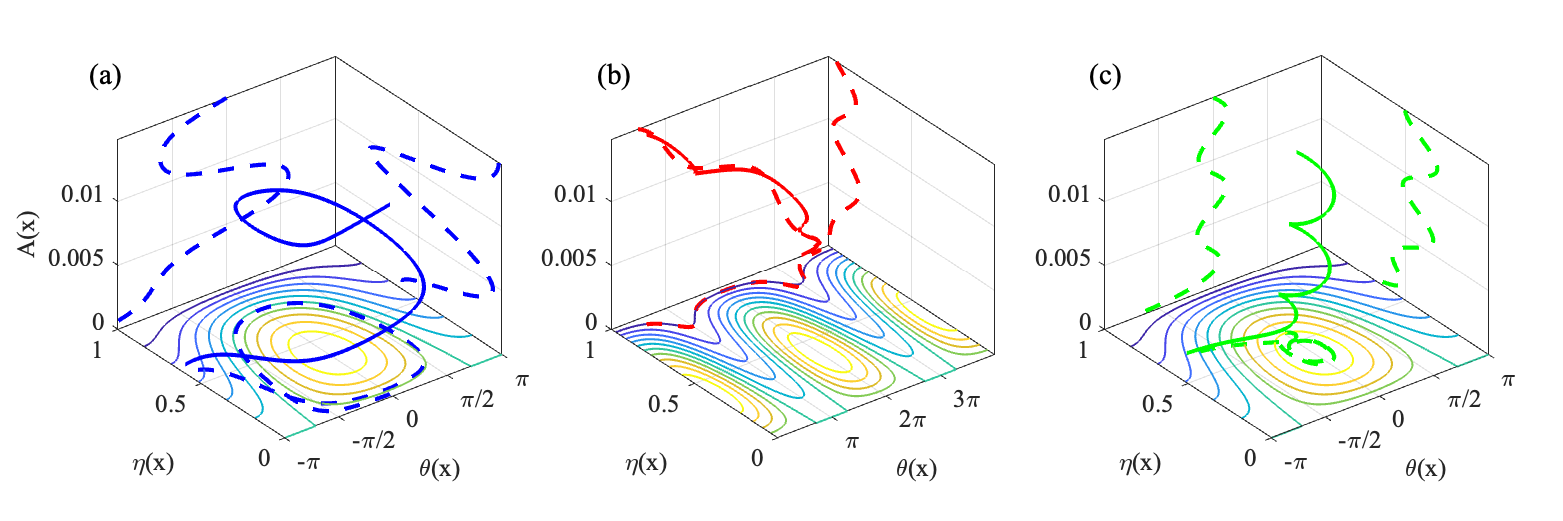}
    \caption{Three indicative phase portraits for a configuration with a carrier $f_a = 1$ Hz, $\epsilon_a = 0.1$ and side band separation $p=0.184.$ The path lines in $(A,\eta,\theta)$-space are shown as solid curves, with different initial conditions $\eta(0), \, \theta(0)$. The undamped phase portraits in the $(\eta,\theta)$ plane are shown in the $(\eta,\theta)$-plane, along with the projections of the damped path curves ($s=9\cdot10^{-6}, \, \gamma=0.002$) which are depicted as dashed curves.}
    \label{fig:Phase portraits uniform damping w projection}
\end{figure}

Some examples of trajectories in the three-dimensional phase space of \eqref{eq:A' uniform damping}, \eqref{eq:eta' uniform damping}--\eqref{eq:theta' uniform damping} are shown in Figure \ref{fig:Phase portraits uniform damping w projection}. An illustrative uniform damping parameter is chosen, and the trajectories are plotted as solid lines starting at different initial conditions $\eta_0,\theta_0$ but with an identical initial value of $A$ (this value is fixed by a choice of the carrier wave frequency $f_a$ and steepness $\epsilon_a,$ as well as the side-band separation $p,$ see Section \ref{sec:Modelling of damped waves}). 

The left panel ((a), blue curves) begins close to a bichromatic wave train $\omega_b,\, \omega_c$, with $\eta(0)=0.02;$ the middle panel ((b), red curves) is a small perturbation of a monochromatic wave train $\omega_a,$ with $\eta(0)=0.95;$ the right panel ((c), green curves) is a trichromatic initial condition $\eta(0)=0.5$ where no mode is initially dominant. The projections of the trajectories onto the planes $(\eta,\theta), \, (A,\theta)$ and $(A,\eta)$ are shown as dashed curves, and the phase portraits of the corresponding conservative system (at $x=0$) with constant $A$ are shown in colour on the $(\eta,\theta)$--plane. We can observe both periodic dynamics in phase $\theta$ and amplitude scale $\eta$, as in panel (b), as well as the breaking of such periodic dynamics in panels (a) and (c).

\subsection{Spatial Benjamin-Feir instability with uniform damping}
\label{ssec:UD Spatial BFI}

The famed Benjamin-Feir instability arises in the undamped system when a carrier wave $\omega_a$ is perturbed by two equally spaced side-bands $\omega_a+p$ and $\omega_a-p.$ In the current setting, we find that the monochromatic carrier wave which makes up the nullcline $\eta = 1$ is unstable if there exists a fixed point of the dynamical system \eqref{eq:eta' uniform damping}--\eqref{eq:theta' uniform damping} thereon.

\begin{figure}
    \centering
    \includegraphics[width=0.85\linewidth]{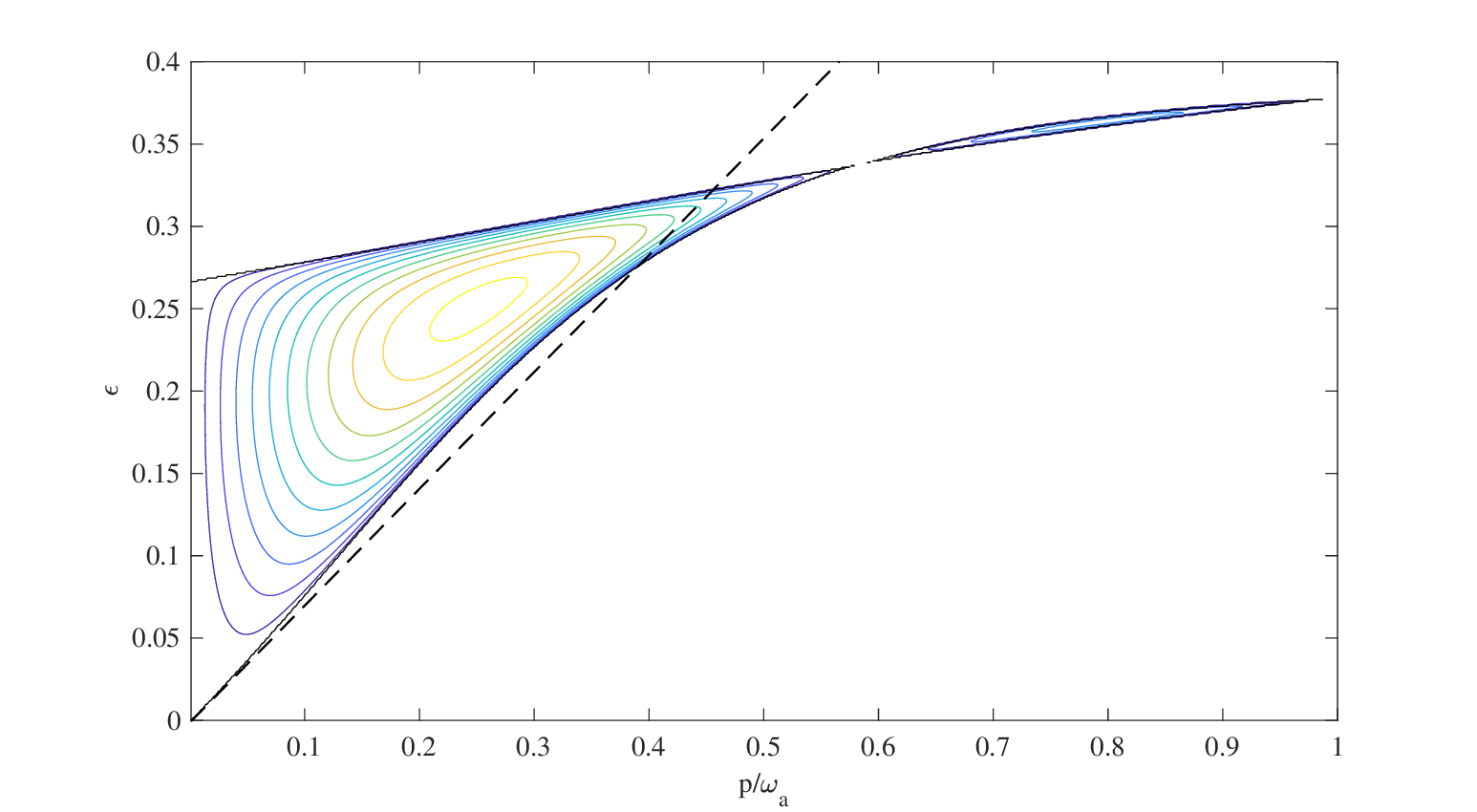}
    \caption{Region of $(\epsilon,p)$ parameter space showing the domain of instability for a carrier wave with frequency $f_a = 1$ Hz. Lighter colours denote a larger linear growth rate. The dashed line shows the comparable instability threshold for the spatial NLS, given by $p/\omega_a < \sqrt{2} \epsilon.$}
    \label{fig:BFI Growth Rates No Damping}
\end{figure}

This means solving 
\begin{equation*}
    \cos(\theta) = -\frac{\Delta + A \Omega_0 + A \Omega_1}{A \Omega_2},
\end{equation*}
which indicates that a necessary and sufficient condition for such a fixed point is that the right-hand side is of magnitude less than or equal to one. The (linear) growth rate of disturbances is given by the eigenvalues of the Jacobi matrix of the undamped Hamiltonian system, i.e.\ $\lambda_{1,2} = \pm \sqrt{A^2 \Omega_2^2 \sin^2 \theta},$ evaluated at the fixed point, and is plotted in Figure \ref{fig:BFI Growth Rates No Damping}. This is the spatial version of the famed temporal instability diagram for Stokes waves. The NLS instability threshold is shown as a dashed line, with all waves to the left being unstable; as expected, this matches the Zakharov equation result for small mode separation $p.$

\begin{figure}[h!]
\centering
\includegraphics[width=0.85\linewidth]{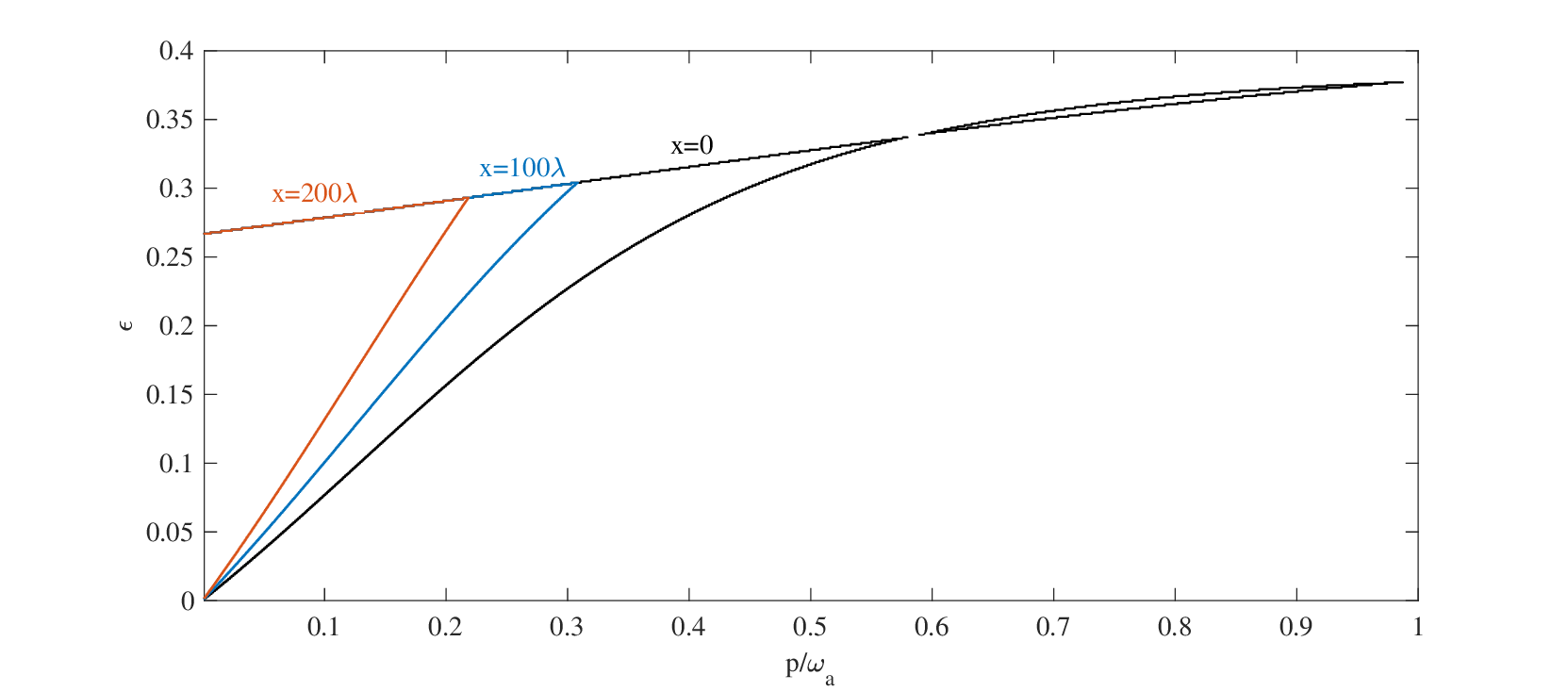}
\caption{Stability domain for a monochromatic wave with $f=1$ Hz in terms of (initial) steepness $\epsilon$ and mode-separation $p$ (black region, see also Figure \ref{fig:BFI Growth Rates No Damping}). The stability region after a propagation distance $100\lambda$ (blue region) and $200 \lambda$ (red region) is shown for uniform damping $s=7\cdot10^{-6}, \, \gamma=1.7\cdot10^{-3}.$}
\label{fig:stabilisation with propagation}
\end{figure}

In the relatively simple uniform damping scenario, the ODE for $A$ is decoupled from the equations governing the energy exchange $\eta$ and dynamic phase $\theta.$ In particular, this means that 
\[ A = A_0 \exp(-2\gamma x)\]
for $A_0 = \ca I_a(0) + \cb I_b(0) + \cc I_c(0).$
Thus waves which are initially unstable will stabilise after a certain propagation distance, depending on the concrete interplay between the damping $\gamma,$ the mode separation $p$ and the initial steepness. This can be observed in Figure \ref{fig:stabilisation with propagation}, which demonstrates how the stability of waves with given $\epsilon,p$ changes with propagation distance in the damped case. It can be observed that for a given mode separation progressively steeper waves remain unstable as the waves propagate into the damped region. Conversely, the diminution of wave steepness apparent in the propagation of damped waves means that disturbances far from the carrier in Fourier space stabilise with propagation distance; for example, all disturbances $p>0.3\omega_a$ have stabilised after 100 carrier wavelengths $\lambda$. Those disturbances with the largest linear growth rates are among the first to stabilise. The consequences of these effects will be particularly notable in our discussion of spectral broadening in Section \ref{sec:spectral broadening}.

\section{A dynamical system for non-uniform damping}
\label{sec:Dynamical system for nonuniform damping}

Non-uniform damping complicates the situation somewhat, and ensures that we must retain an equation for the side-band energy fraction $\alpha.$ Introducing the change of variables from $I_i$ to $A, \, \eta, \alpha$ as in Section \ref{sec:A dynamical system for uniform damping}, we find that our equation for the evolution of the wave--action becomes 
\begin{equation*}
    \frac{dA}{dx} = -2( c_{g,a}\gamma_aI_a + c_{g,b}\gamma_bI_b + c_{g,c}\gamma_cI_c).
\end{equation*}
Without the assumption of equal side-band energy fraction we must write 
\begin{equation*}
    \frac{d \alpha}{dx} = -2 \left(\cb \gamma_b I_b - \cc \gamma_c I_c\right).
\end{equation*}
The reformulation in terms of an energy-exchange parameter $\eta$ must likewise be altered to
\begin{align} \label{eq:doublevar1}
    I_a &= \frac{A}{\ca} \eta, \\ \label{eq:doublevar2}
    I_b &= \frac{A}{2\cb}(1-\eta+\alpha), \\ \label{eq:doublevar3}
    I_b &= \frac{A}{2\cc}(1-\eta-\alpha).
\end{align}
Then 
\begin{equation} \label{eq:alpha' nonunif}
    \frac{d \alpha}{dx} = \alpha^2 (\gamma_b - \gamma_c) + \alpha \eta (2 \gamma_a - \gamma_b - \gamma_c ) + (\eta-1) (\gamma_b - \gamma_c),
\end{equation}
and
\begin{equation} \label{eq:A' nonunif}
    \frac{dA}{dx} = -2 A \eta (\gamma_a - \frac{1}{2}(\gamma_b + \gamma_c)) - A (\gamma_b (1+\alpha) + \gamma_c (1-\alpha)).
\end{equation}
In terms of this we write the evolution of $\eta(x)$ as 
\begin{equation} \label{eq:eta' nonunif}
    \frac{d \eta}{dx} = 2 \eta^2 \left( \gamma_a - \frac{1}{2} (\gamma_b + \gamma_c ) \right) + \eta \left(\gamma_b (1+\alpha) + \gamma_c (1-\alpha) \right) - 2 \gamma_a \eta - A \Omega_2 \eta \sqrt{(1-\eta)^2 - \alpha^2} \sin(\theta).
\end{equation}

The evolution of the dynamic phase can be written
\begin{equation} \label{eq:theta' nonunif}
    \frac{d\theta}{dx} = \Delta + A \Omega_0 + A \Omega_1 \eta - A \Omega_2 \left[ \frac{(1-\eta)(1-2\eta)-\alpha^2}{\sqrt{(1-\eta)^2-\alpha^2}} \cos(\theta) \right],
\end{equation}
where the generalised expressions for $\Omega_0, \, \Omega_1$ are
\begin{align*}
    \Omega_0 &= 2 \left[ \frac{1}{4} \left( \bar{T}_c (1-\alpha) + \bar{T}_b(1+\alpha) \right) + \bar{T}_{bc} - \left( \bar{T}_{ac}(1-\alpha) + \bar{T}_{ab}(1+\alpha) \right) \right],\\
    \Omega_1 &= 4(\bar{T}_{ab} + \bar{T}_{ac}) - 2\bar{T}_{bc} - (2\bar{T}_a + \frac{1}{2}(\bar{T}_b + \bar{T}_c)).
\end{align*}
Note that these reduce to the corresponding expressions in Section \ref{sec:A dynamical system for uniform damping} for $\alpha=0,$ and that $\Omega_2$ is unchanged.

The phase space of equations \eqref{eq:alpha' nonunif}--\eqref{eq:theta' nonunif} is now four dimensional, and occupies 
\[ \{ (A,\alpha,\eta,\theta) \mid A \in \mathbb{R}_{\geq 0}, \, \alpha \in \mathbb{R}, \, 0 \leq \eta \leq  \min(1-\alpha,1+\alpha), \, \theta \in [-\pi,\pi] \}.\]
In particular, while $\eta=0$ corresponds to the presence of modes $\omega_b$ and $\omega_c$ only (i.e.\ a bichromatic sea), for nonzero $\alpha$ there is no monochromatic sea consisting of only wave $\omega_a$ (since $1-\eta+\alpha$ and $1-\eta-\alpha$ are not simultaneously zero for any value of $\eta$). It is still possible to study the Benjamin-Feir instability with an initial side-band imbalance s.t.\ $\alpha\neq0,$ but we note that the energy transfer from carrier to side-bands captured by \eqref{eq:doublevar1}--\eqref{eq:doublevar3} remains symmetric. The more interesting scenario, which we explore below, is how nonuniform damping generates side-band imbalances, and how it differs from the simpler, uniformly damped dynamics.

\section{Modelling of damped waves with applications to sea--ice}
\label{sec:Modelling of damped waves}

To understand the evolution of surface gravity waves damped due to the presence of sea ice we must relate the complex amplitudes and the conserved quantity $A$ to physical properties of the waves. For a single wave, the free surface elevation \eqref{eq:Free Surface} can be written as 
\begin{equation*}
    \zeta(x,t) = \frac{1}{\pi} \sqrt{\frac{\omega_a}{2g}} |B_a| \cos(\xi_a + \phi_a),
\end{equation*}
where $\xi_a = k_a x - \omega_a t$ and $\phi_a$ is the phase of mode $k_a.$ The relation $A=\ca |B_a|^2$ allows us to rewrite
\begin{equation*}
    A = \frac{a_a^2 \pi^2 g}{k_a} = \frac{\epsilon_a^2 \pi^2 g}{k_a^{3}},
\end{equation*}
where $a_a$ is the physical amplitude of the wave, and $\epsilon_a = a_a k_a$ is the wave steepness. 

For the three-mode system describing Benjamin-Feir instability, we can write the following amplitude function
\begin{equation*}
A(x,t) = \frac{1}{\pi} \left( \sqrt{\frac{\omega_a}{2g}} B_a e^{i(k_a x - \omega_a t)} +  \sqrt{\frac{\omega_b}{2g}} B_b e^{i(k_b x - \omega_b t)} + \sqrt{\frac{\omega_c}{2g}} B_c e^{i(k_c x - \omega_c t)} \right).
\end{equation*}
Then the free surface elevation $\zeta(x,t)$ is obtained from $\zeta(x,t)=\mathfrak{Re}[A(x,t)]$, while the free-surface envelope is $|A(x,t)|$.

It is instructive to explore some cases through the lens of the three equivalent formulations, using the reduced variables introduced in \eqref{eq:alpha' nonunif}--\eqref{eq:theta' nonunif} and ultimately obtaining therefrom the free surface envelope. The point of departure is the choice of an initially unstable carrier wave, which we take to be $f=1$ Hz and have a steepness of $\epsilon=0.15.$ (In principle an initial, stable carrier could be chosen as shown in Figure \ref{fig:BFI Growth Rates No Damping}, either with large steepness above the restabilisation threshold \cite{Yuen1982} or very low steepness such that instability is confined to very small side-band separation and growth rates are low. In both cases the waves are nearly linear except for a dispersion correction \cite{Stuhlmeier2019}, and their behaviour is dominated by the decay in energy \eqref{eq:A' uniform damping}, making these cases less interesting.)

Having chosen a carrier, we select side bands which form the small initial disturbance. To provide contrast, one pair of side-bands is taken well within the unstable region, with mode-separation $p=0.4;$ the second pair, with lower linear growth rate is selected at $p=1$ (for reference the instability domain in Figure \ref{fig:BFI Growth Rates No Damping} at $\epsilon=0.15$ extends to $p\approx 1.2$). An equidistribution of side-band energy is originally imposed by setting $\eta(0)=0.98$ and $\alpha(0)=0,$ and the initial value of the dynamic phase is set equal to $\theta(0)=\pi/2.$ These choices together determine the initial trajectory. An illustrative moderate damping parameter of $s=7\times10^{-6}$ is employed, and frequency dependent damping implemented as $\gamma_i=s \omega_i^3$ as described in \eqref{eq:damping-form}.

\begin{figure}[h]
    \centering
    \includegraphics[width=\linewidth]{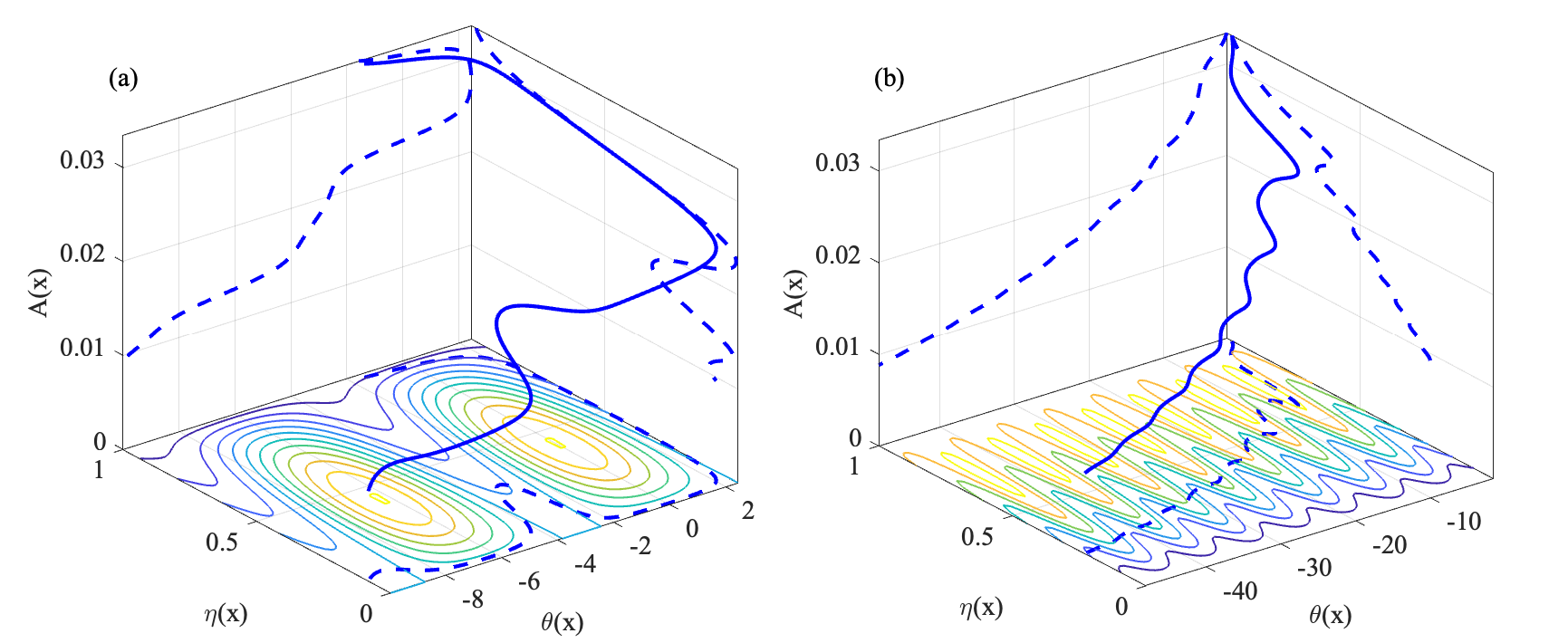}
    \caption{Phase portraits for two cases with intermediate, non-uniform damping. In both cases the carrier is initiated at $f=1$ Hz, damping coefficient $s=7\times10^{-6},$ $\epsilon=0.15$ and the initial conditions are $\eta(0)=0.98,$ $\theta(0)=\pi/2.$ Panel (a) shows the evolution over $x=400$ m (256 $\lambda_a$) of the initially unstable carrier to perturbations with $p=0.4,$ while panel (b) shows the evolution when the perturbations are located at $p=1.$}
    \label{fig:two-cases-phase-portraits}
\end{figure}

The evolution in three-dimensional phase space $(A,\eta,\theta)$ of these two configurations is shown in Figure \ref{fig:two-cases-phase-portraits}. The initial condition is the same in each case, yet the interaction between the modes is modulated by the side-band separation, and thus governed initially by the Hamiltonian of the conservative system whose phase portrait is shown projected onto the $(\eta,\theta)$-plane. Note that the phase portraits of the conservative system depend on $A$ as well as $p$, and are therefore not identical in right and left panels.

After nearly 256 carrier wavelengths (corresponding to 400 m), the total energy $A(x)$ has decreased to one third of its initial value in both cases. The more unstable case shown on the left (with $p=0.4$) shows the initial dominance of nonlinear interaction, as the trajectory winds around the centre located at $\theta=0$. By contrast, the less unstable case in the right panel (with $p=1$) is dominated by dissipation. The trajectory traverses the separatrix of the conservative system almost immediately, and thereafter only small oscillations indicating energy exchange among the modes are visible.

\begin{figure}[h]
    \centering
    \includegraphics[width=\linewidth]{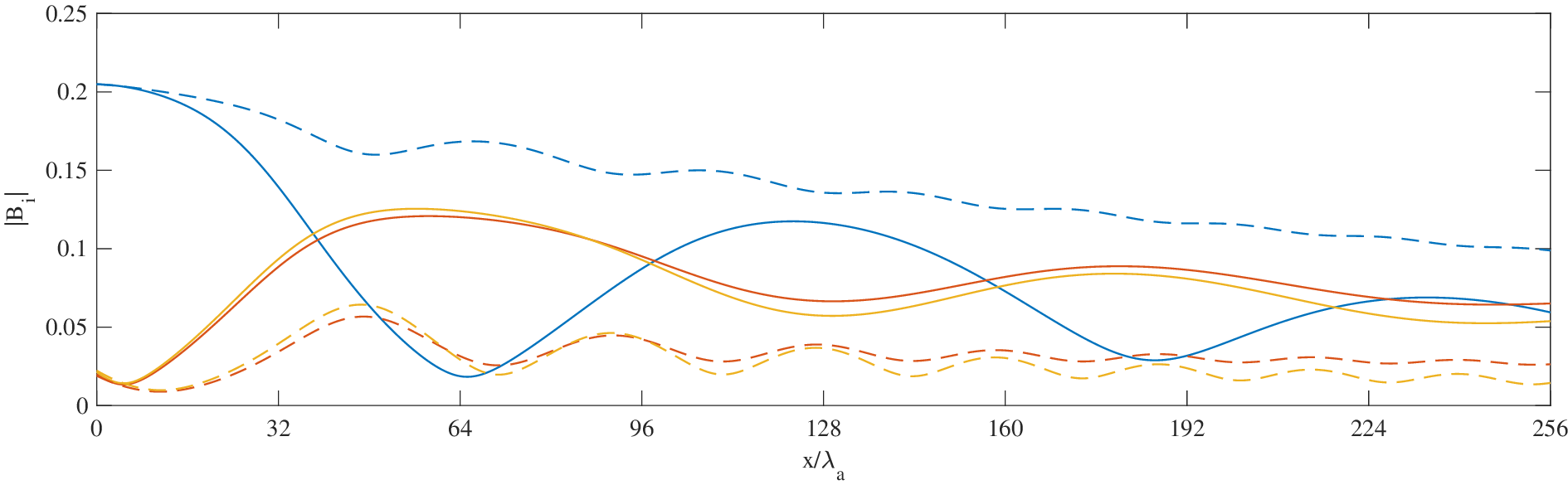}
    \caption{Evolution of the complex amplitudes shown in Figure \ref{fig:two-cases-phase-portraits}. Solid curves correspond to $p=0.4,$ while dashed curves correspond to $p=1.$ The carrier at $f=1$ Hz is shown in blue, the two side-bands in red and yellow.}
    \label{fig:two-cases-modal-amplitudes}
\end{figure}

The consequences of this dissipative interaction are visualised for the complex magnitudes $|B_i|$ individually in Figure \ref{fig:two-cases-modal-amplitudes}. The solid curves depict the stronger interaction ($p=0.4$) while the dashed curves depict the weaker interaction ($p=1$). The interplay between the carrier $|B_a|$ in blue and the two side-bands $|B_b|, \, |B_c|$ in yellow and red is also traced out in the projection onto the $(A,\eta)$-plane in Figure \ref{fig:two-cases-phase-portraits}, where $\eta=1$ indicates only the carrier is present (monochromatic sea), and $\eta=0$ means only the side-bands are (bichromatic sea).

\begin{figure}[h]
    \centering
    \includegraphics[width=\linewidth]{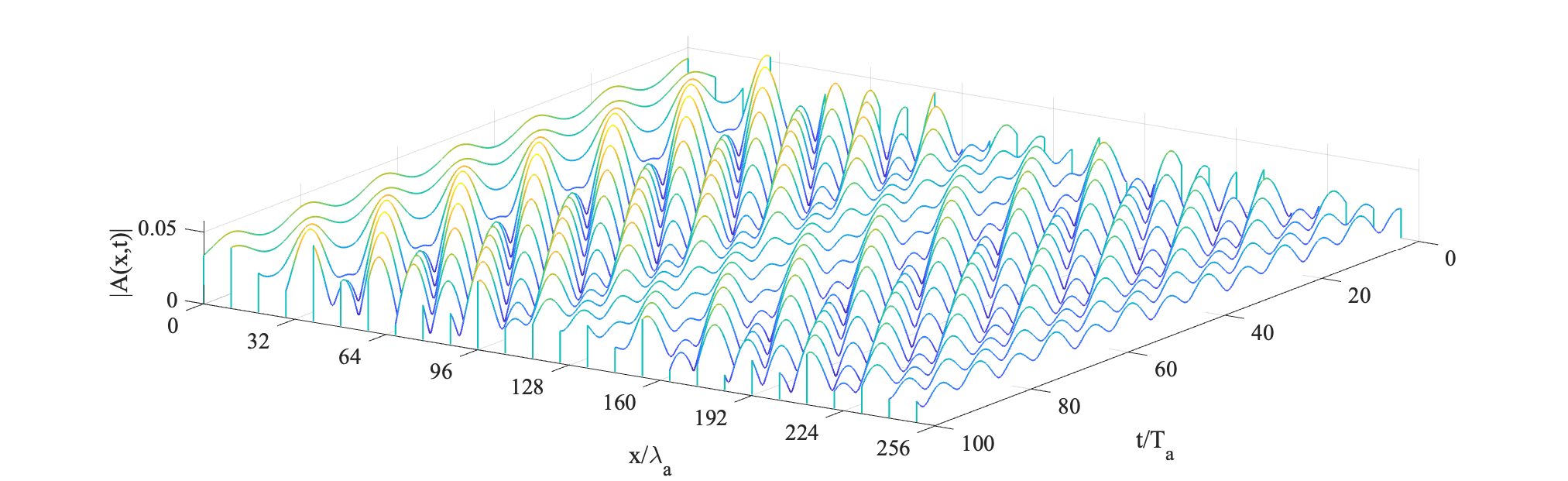}
    \caption{Free surface envelope for the case $p=0.4$ shown in Figures \ref{fig:two-cases-phase-portraits}--\ref{fig:two-cases-modal-amplitudes}.}
    \label{fig:two-cases-case-1-free-surface}
\end{figure}

Finally, in Figure \ref{fig:two-cases-case-1-free-surface} we can observe how the modulational instability combined with damping influences the free surface envelope $|A(x,t)|$ in space and time. For clarity only the case $p=0.4$ is presented, and we can clearly distinguish the two cycles of energy exchange among the modes seen in Figure \ref{fig:two-cases-modal-amplitudes}. It is helpful to recall that a monochromatic wave has constant envelope, so that the absence of side-band energy observed at $x=0$ and $x=128 \lambda_a$ manifests in a flattening of the envelope in Figure \ref{fig:two-cases-case-1-free-surface}. We also note that the waves are periodic in time -- at a given spatial propagation distance we find a surface envelope time-series as given by one of the sections shown in the figure.

\subsection{Uniform versus nonuniform damping}
\label{ssec:Uniform vs nonuniform damping}

One advantage of the reformulation in terms of energy scale $\eta$ and dynamic phase $\theta$ is that it clearly demonstrates that the full, nonlinear energy exchange in the modulational instability is naturally symmetric. Rewriting the mode amplitudes in terms of energy scale $\eta$ and side-band difference $\alpha$ as in \eqref{eq:singlevar1}--\eqref{eq:singlevar3} or \eqref{eq:doublevar1}--\eqref{eq:doublevar3} makes clear that when the carrier loses energy, that energy is distributed equally among the higher and lower side-bands.

However, frequency dependent damping can break this symmetry. Indeed, when the damping is nonuniform, even initially equal side-bands, such as those which initialise the classical Benjamin-Feir instability, eventually develop an imbalance, as previously shown by Alberello et al \cite{Alberello2023} using a dissipative NLS framework. The most striking examples of how uniform and nonuniform damping differ can be found in a synthetic situation in which the Fourier amplitudes of all three modes $I_a, \, I_b$ and $I_c$ are initially taken to be equal. If such a triad is unstable, energy exchange means that mode $I_a$ will transfer energy to modes $I_b$ and $I_c$ and vice versa, in a recurrent fashion dependent on the initial value of the phase \cite{Andrade2023}. 

\begin{figure}[ht]
\centering
\includegraphics[width=\linewidth]{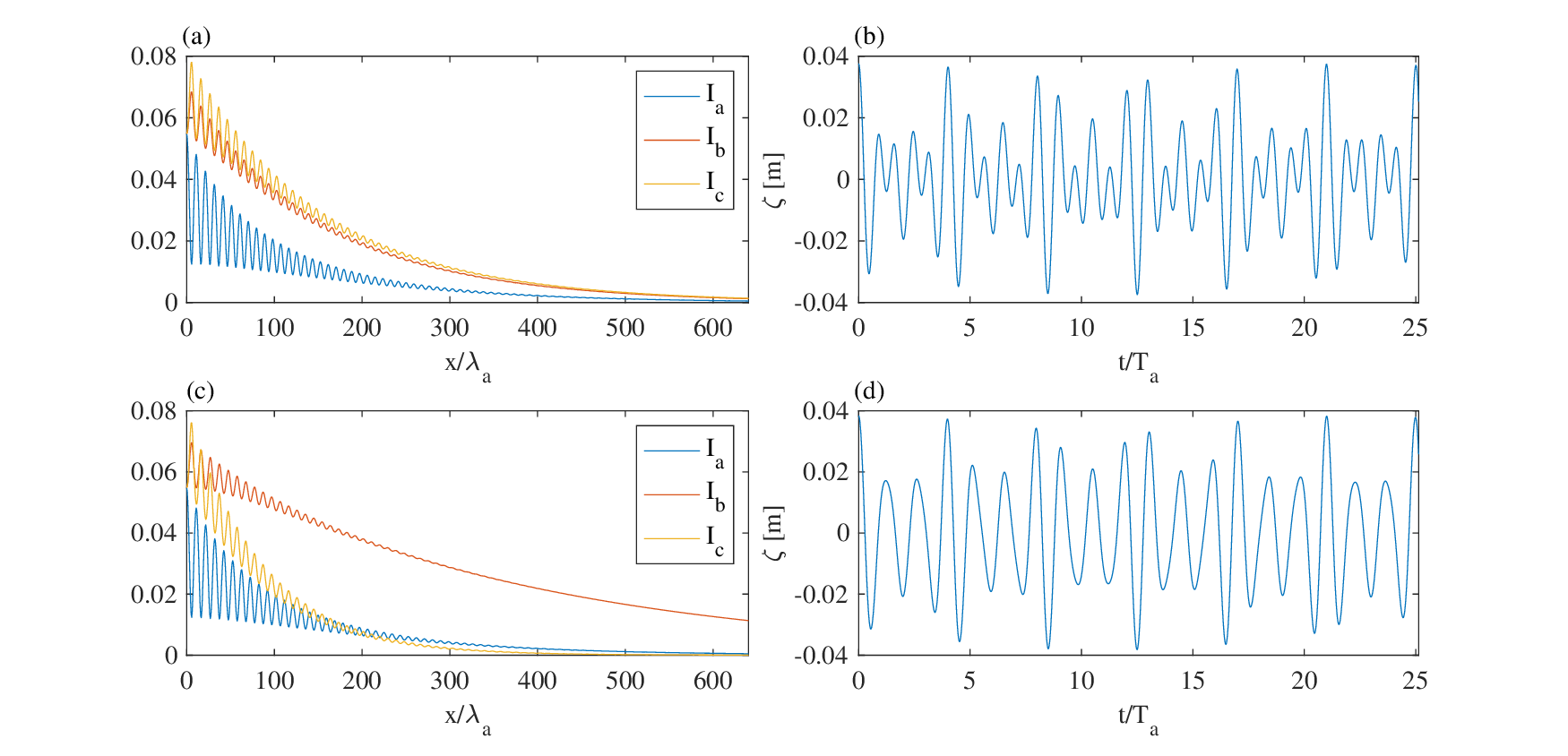}
\caption{Case initialised with $f$=1 Hz, $\epsilon=0.3, \, p = 1.5, \alpha\approx 0.1622$ and $\eta(0)\approx0.3204, \theta(0)=\pi,$ where $\alpha$ and $\eta$ are chosen so that initially $I_a(0)=I_b(0)=I_c(0).$ (Panels (a) \& (b)) Uniform damping with $\gamma=8\times10^{-6} \omega_a^3.$ Panel (a) shows the modal amplitudes $I_i$ with space; panel (b) shows the free surface at $x=800$ m (512 $\lambda_a$) as a function of time. (Panels (c) \& (d)) Nonuniform damping with $\gamma_i=8\times10^{-6} \omega_i^3.$ Panel (c) shows the modal amplitudes $I_i$ with space; panel (d) shows the free surface at 512 carrier wavelengths or $x=800$ m. This illustrates clearly that non/uniform damping can change the dominant wave and the character of the wave field at a given spatial location.}
\label{fig:Uniform-vs-nonuniform-damping}
\end{figure}

With uniform damping, modes $I_b$ and $I_c$ decay at the same rate, while nonuniform damping with positive exponent (as in our case) induces the shorter wave to decay faster. This situation is depicted in Figure \ref{fig:Uniform-vs-nonuniform-damping}, where the top two panels show uniform damping in the amplitude spectrum (a) and the free surface at a fixed spatial location (b), while the bottom panels depict the same for nonuniform damping proportional to $\omega_i^3$. It is seen that nonuniform damping can dramatically change which modes are dominant after a given propagation distance.

\subsection{Spectral broadening, chaotisation and the damped Benjamin-Feir instability}
\label{sec:spectral broadening}

The cases considered hitherto have been restricted to three modes, and are thus amenable to an analytical description. The textbook modulational instability only considers the initial exponential growth arising from linear stability theory, while our description is able to capture both the instability and the subsequent behaviour in the presence of damping.  However, experimental and theoretical work points to the fact that the Benjamin-Feir instability -- either with or without damping -- may in many cases entrain higher harmonics, which marks a departure from our description\cite{Yuen1982}.

\begin{figure}[h]
\centering
\includegraphics[width=\linewidth]{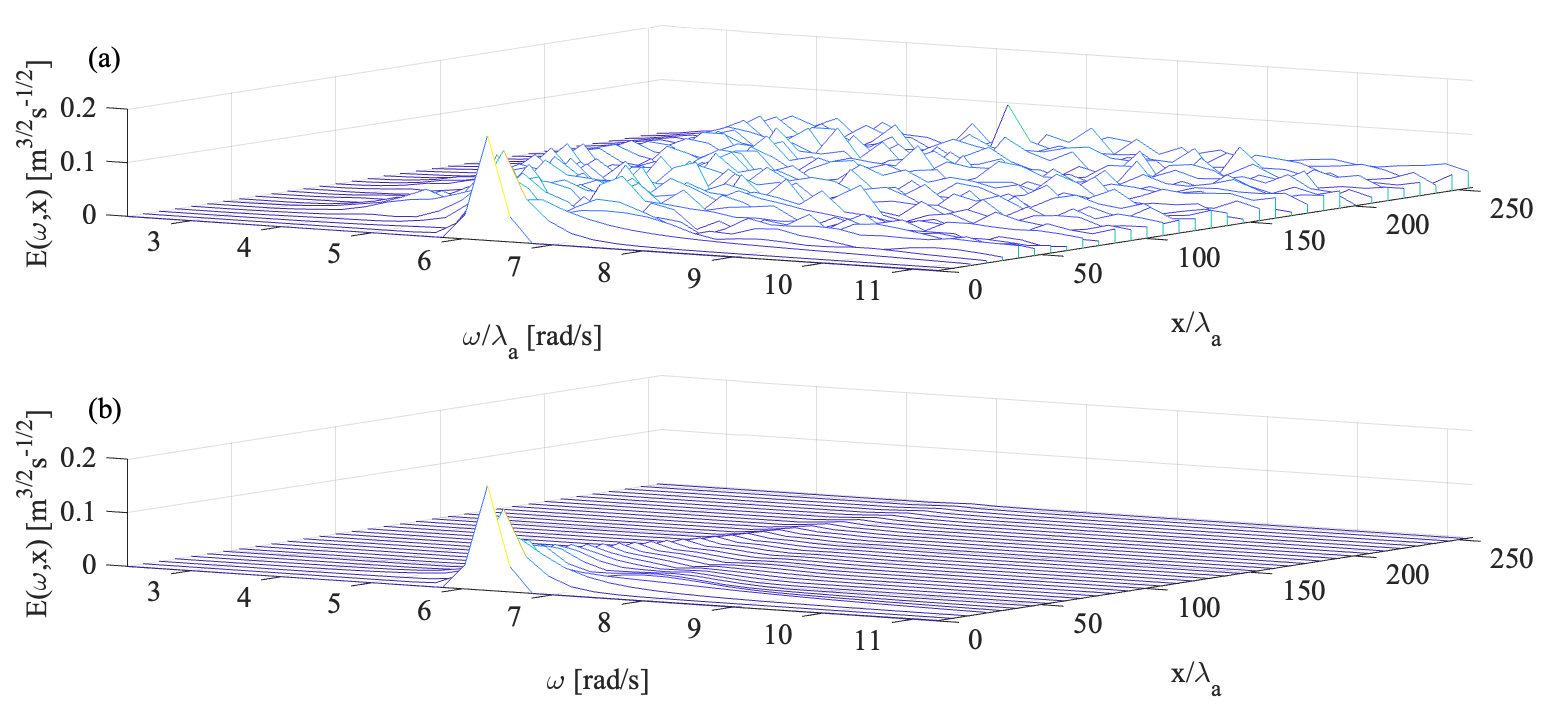}\\
\caption{Evolution of an initial unstable wave, with carrier frequency $f=1$ Hz, $p=0.25, \eta=0.9, \, \theta=0, \, \epsilon=0.15.$ The initial triad is initialised among 41 equally spaced Fourier modes. Panel (a):  depicts the evolution obtained from numerical integration of the spatial Zakharov equation without damping. Panel (b): when a frequency dependent damping with $s=5\times10^{-5}$ is incorporated this arrests the spectral broadening completely, and leads only to a spectral downshift.}
\label{fig:Multimode-w-damping}
\end{figure}

To this end it is instructive to consider the effect of including further modes, which makes it possible to obtain some insight into the evolution of an initially unstable wave-train in more realistic conditions. One such case is shown in Figure \ref{fig:Multimode-w-damping}, where the same initial configuration is evolved by solving a spatial Zakharov equation with 41 modes with and without frequency-dependent damping. The initial spectrum $E(\omega,x=0)$ consists of the three Fourier modes $I_a, \, I_b$ and $I_c$ in both cases. In the top panel (a) we observe the evolution of the spectrum in space without damping, which shows a characteristic broadening. The bottom panel (b), in contrast, shows the effects of strong frequency-dependent damping (with $s=5\cdot10^{-5}$), which inhibits the spectral broadening and leads to a distinct downshift in the peak frequency.

\begin{figure}[h]
\centering
\includegraphics[width=\linewidth]{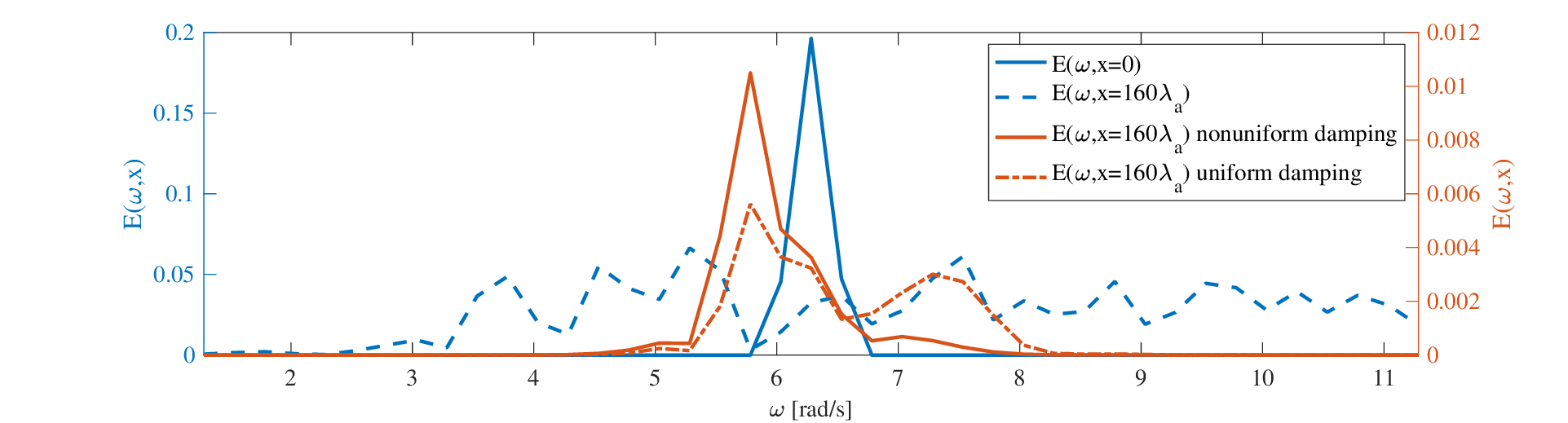}\\
\caption{Fourier amplitude spectra for the configuration studied in Figure \ref{fig:Multimode-w-damping}. The solid blue curve shows the initial, 3-mode spectrum at $x=0$. The undamped spectrum (blue dashed curve; on left scale) shows significant spectral broadening after a propagation distance of $x=160$ carrier wavelengths. By contrast, the damped spectra (solid curve: frequency-dependent damping, dash-dotted curve: uniform damping; both on right scale) exhibit a downshift with nearly negligible broadening.}
\label{fig:Multimode-w-damping-spectral-evol}
\label{fig:Multimode-w-damping-unif-nonunif-spectral-evol}
\end{figure}

This spectral downshift is illustrated more clearly in Figure \ref{fig:Multimode-w-damping-spectral-evol}, which shows sections through Figure \ref{fig:Multimode-w-damping} at two points in the evolution. The initial tri-modal spectrum at $x=0$ is shown as a solid blue curve, and the undamped spectrum after $160$ peak wavelengths of propagation distance is shown as a blue, dashed curve. This clearly shows how energy has spread among neighbouring Fourier modes. The spectrum with nonuniform damping (as in Figure \ref{fig:Multimode-w-damping}, bottom panel) is shown at $x=160\lambda_a$ as a solid red curve, and makes clear the inhibition of spectral broadening. For comparison the same spectrum is propagated with uniform damping -- using the frequency of the carrier $\omega_a$ to determine the uniform damping parameter $\gamma$ -- and shown as a red, dash-dotted curve. 

In fact, the interplay between energy exchange (which itself can lead to spectral broadening and a frequency downshift in the absence of damping \cite{Janssen2004}) and damping is complex, as demonstrated in Figure \ref{fig:two-cases-phase-portraits}. While frequency-dependent damping will engender a narrower spectrum (as can be seen by comparing the solid and dash-dotted red curves in Figure \ref{fig:Multimode-w-damping-spectral-evol}) due to stronger damping at the higher frequencies, the spectral evolution is also dependent on the initial phases and the carrier amplitude. It is also important to emphasise that the spectra here considered are amplitude spectra rather than energy spectra, i.e.\ no phase averaging has taken place.

\section{Discussion}
\label{sec:Discussion}

Recent years have seen a great deal of interest in wave interactions in the presence of damping, particularly with applications to wave propagation in sea ice. Many previous studies have focused on the use of partial differential equations such as the NLS or the Dysthe equation, which are inherently restricted to narrow bandwidth as shown in the derivation of the damped spatial NLS equation in Section \ref{ssec:Derivation of dNLS}. Our first methodological novelty thus consists in relaxing this assumption, and employing the Zakharov equation derived directly from the cubic reduced Hamiltonian formulation, without further restrictions on the spectral width. The cubic nonlinearity is sufficient to capture the leading-order energy exchange which gives rise to the modulational instability, and thus provides an ideal point of departure.

While the Zakharov formulation has been successfully exploited in water wave modelling for theoretical and practical purposes for the past three decades, the propagation of waves into ice-covered waters requires a reformulation of this problem. Physically we envision undamped, periodic waves from the open sea encountering an area of sea ice (or any other non uniformly dissipative medium) of fixed spatial extent. As these waves propagate into the sea ice, the effects of damping are felt with propagation distance rather than time. This means that a spatial Zakharov formulation with damping is required, which is developed in Section \ref{ssec:Spatial ZE & Damping}. The spatial Zakharov formulation is also appropriate for flume experiments, and several papers testing its applicability in the absence of damping exist in that context \cite{Galvagno2021,Shemer2017}.

Modulational instability is triggered when a wave train of given frequency $\omega_a$ is perturbed by the introduction of a pair of low-amplitude satellites having a higher and a lower frequency, $\omega_b$ and $\omega_c$, say. This is a type of near-resonant interaction where $2\omega_a = \omega_b + \omega_c,$ and where we denote the separation between the carrier $\omega_a$  and side-band harmonics (or Fourier modes) $\omega_b, \, \omega_c$ by $p.$ Whether a given carrier is unstable depends on its steepness or energy, as well as the side-band separation $p,$ and the consequences of this instability have been extensively studied in the temporal evolution scenario. Rather surprisingly, few studies of spatial evolution within the Zakharov equation context exist, with experimental work by Shemer \& Chernyshova \cite{Shemer2017} and theoretical work by Kachulin et al \cite{Kachulin2019} and Dyachenko et al \cite{Dyachenko2017} being notable exceptions.

In fact, it is possible to make an analytical study of modulational instability when only the carrier and the two side-bands are considered. This was undertaken in the context of the NLS by Capellini \& Trillo \cite{Cappellini1991}, and more recently by Andrade \& Stuhlmeier \cite{Andrade2023,Andrade2023instability} using the Zakharov formulation. The key to this reformulation of the problem is the identification of a single \textit{dynamic phase variable}, which combines the individual modal phases according to the resonance condition. The individual phases are largely irrelevant for the dynamics, and this simple identification immediately reduces the dimension of the system by two. In the conservative case the conservation of energy (Hamiltonian), wave action, and momentum can be used to reduce the modulational instability to a planar dynamical system in terms of dynamic phase and energy scale.

This reduction is still useful when the system is nonconservative due to the effects of damping; indeed, the reformulated system gives insight into the energy exchange processes, while stripping out physically irrelevant information. Thus, while the total energy of the system decreases steadily, at each energy level the trajectories in the three-dimensional phase space are directed by the associated conservative system, as shown in Section \ref{sec:A dynamical system for uniform damping}. Damping means that separatrices of the conservative system can be crossed, and fixed points disappear. A competition sets in between the effects of damping and those of energy exchange, with the former dominating for small initial growth rates, while the latter dominates when the modulational instability growth rate is high. 

Damping may also have the effect of shifting the dominant wave, particularly when the damping depends on the wave frequency, as is the case for propagation in sea-ice. The consequences of frequency dependent damping can thus be observed in the modal amplitudes as well as in the appearance of the free surface, which we demonstrate in Section \ref{ssec:Uniform vs nonuniform damping}. It is also possible to allow many more modes to interact, and thus study how the modulational instability gives rise to spectral broadening. While this case is no longer amenable to analytical insight, we can numerically integrate the damped, spatial Zakharov equation starting with an initially unstable carrier and two small side-bands. We readily observe that the effect of damping can induce a spectral downshift, as observed experimentally for waves propagating in sea ice  \cite{Waseda2022}, as well as inhibit spectral broadening which occurs in the undamped configuration. Indeed, after hundreds of wavelengths the damped amplitude spectrum remains remarkably narrow and confined, a scenario which can be observed in when waves propagate into sea ice.

While we have sought to provide an elegant, analytical description of the damped modulational instability in space, it may also be interesting to consider the statistics of damped waves and the propagation of wave energy spectra (with random phases) in the presence of frequency-dependent damping using the Zakharov formulation, thus extending previous work by Alberello \& P\u{a}ra\u{u} \cite{Alberello2022}. Further avenues of study might also explore the combined effects of wind-forcing and damping on the Zakharov formulation, and compare this with recent work by Armaroli et al \cite{Armaroli2018} using the Dysthe equation formulation. Extensions to other physical scenarios such as internal waves \cite{Ivanov2022} or capillary-gravity waves \cite{Krasitskii1994,Martin:2017} could likewise present interesting avenues for future work.

\section*{Acknowledgements}
RS and CH gratefully acknowledge the support of EPSRC Grant EP/V012770/1.
AA and EP gratefully acknowledge the support of EPSRC Grant EP/Y02012X/1.

\end{document}